\documentclass[a4paper,twoside]{article}
\usepackage{amssymb}
\usepackage{amsmath}
\usepackage{multirow}
\usepackage{url}
\usepackage{color}

\newcommand{\affil}[1]{$^{\rm #1}$}
\usepackage[authoryear]{natbib}
\bibpunct{(}{)}{;}{a}{}{,}
\usepackage{graphicx}
\date{} 
\title{\large\bf\flushleft A Distributed GPU-based Framework for real-time 3D Volume Rendering of Large Astronomical Data Cubes}
\author{\parbox{\textwidth}{\flushleft
\vspace{-0.5cm}
{\it A.H. Hassan\affil{A,C}, C.J. Fluke\affil{A}, and D.G. Barnes\affil{B}}\\
\vspace{0.4cm}
{\small \affil{A}\,Centre for Astrophysics \& Supercomputing, Swinburne University
of Technology, Hawthorn, Victoria, Australia}\\
{\small \affil{B}\,Monash e-Research Centre, Monash University, Clayton, VIC 3800, Australia}\\
{\small \affil{C}\,Email: ahassan@swin.edu.au}}}
\begin{document}
\twocolumn[
\begin{changemargin}{.8cm}{.5cm}
\begin{minipage}{.9\textwidth}
\vspace{-1cm}
\maketitle
%
%
\small{\bf Abstract:}
We present a framework to interactively volume-render three-dimensional data cubes 
using distributed ray-casting and volume bricking over a cluster of workstations powered by one or more 
graphics processing units (GPUs) and a multi-core CPU. The main design target for this framework is to provide an in-core visualization solution able to provide three-dimensional interactive views of terabyte-sized data cubes. We tested the presented framework using a computing cluster comprising 64 nodes with a total of 128 GPUs. The framework proved to be scalable to render a 204 GB data cube with an average of 30 frames per second. Our performance analyses also compare between using NVIDIA Tesla 1060 and 2050 GPU architectures and the effect of increasing the visualization output resolution on the rendering performance.
Although our initial focus, and the examples presented in this work, is volume rendering of 
spectral data cubes from radio astronomy, we contend that our approach has applicability to 
other disciplines where  close to real-time volume rendering of terabyte-order 3D data sets 
is a requirement.


\medskip
\medskip
\end{minipage}
\end{changemargin}
]
\small

\section{Introduction}
\label{sc:Introduction}

Radio astronomy is entering a new data-rich era.  
Upcoming facilities such as  the Australian Square Kilometre Array Pathfinder \citep[ASKAP;][]{johnston:2008}, MeerKAT \citep{booth:2009}, the Low Frequency Array \citep[LOFAR;][]{rottgering:2003}, and ultimately the Square Kilometre Array (SKA)\footnote{\url{http://www.skatelescope.org/}} will enable astronomers to observe the radio universe at an unprecedented spatial and frequency resolution. Unfortunately, handling the data from these facilities,
expected to be of terabyte order for individual observations, will pose a significant challenge for current astronomical data analysis and visualization tools (e.g. DS9\footnote{\url{http://hea-www.harvard.edu/RD/ds9/}} and CASA\footnote{\url{http://casa.nrao.edu/}}). Such data volumes are orders of magnitude larger than astronomers, and existing astronomy software, are accustomed to dealing with.  

Perhaps as never before, opportunities exist for approaches based on advances in computing hardware and a wider 
adoption of techniques centred on computer science and scientific computing to 
overcome these challenges.  Of particular interest is the availability
of graphics processing units (GPUs) as low-cost, highly parallel, streaming
co-processors.  These allow implementation of specific algorithms for
visualization and analysis that may not have been (computationally) 
practical for CPU-only distributed systems.

\subsection{Spectral data cube visualisation}

Today, most radio astronomers still rely on two-dimensional (2D) techniques to visualize 
spectral data cubes that are inherently three-dimensional (3D):  two spatial axes and
one frequency or velocity axis.
These 2D techniques display data cubes as a sequence of colour-coded (or flooded-contoured) frames at a reasonable speed to enable astronomers to identify the relationships between these frames, and detect the main features of signal and noise therein. These techniques, which usually rely on holding an entire data cube in CPU memory, are not feasible with the upcoming dataset sizes. 

Wider adoption of 3D visualization techniques has been proposed as an alternative to 2D techniques, particularly for obtaining global views. Spectral data cubes are often characterised by a lack of well defined surfaces and the presence of significant features (i.e. sources) with either
a spatial or spectral extent near or 
below the noise level.  These data characteristics limit the usage of surface 
rendering techniques [e.g iso-surfaces; \citet{beeson:2003} and \citet{hassan:2010a}], 
the adoption of multi-resolution algorithms, or general purpose data visualization tools. For a detailed review of the previous work in spectral data cube visualization, we refer 
the reader to \citet{hassan:2010b}.  

In terms of providing global views of data, volume rendering is the most promising 
candidate, particularly in cases where clear feature segmentation cannot be 
done \citep{beeson:2003,gooch:1995,oosterloo:1995}. Volume rendering is the process of generating a colour coded 2D projection of the 3D data on a user controlled viewport.

Existing astronomy volume rendering implementations have addressed the special characteristics of astronomical data including the selection of transfer functions \citep{gooch:1995,oosterloo:1996}, effective handling of adaptive grids and different data resolutions \citep{kaehler:2006,nadeau:2001,magnor:2005}, and addressing the large data size problem \citep{beeson:2003}. Some of these implementations were directed toward rendering a specific dataset \citep{kaehler:2006,nadeau:2001,magnor:2005}, while others demonstrated the usage of their tool with different datasets, or provide their tool to the public domain. Throughout these papers we can find two main implementation trends. The first trend customizes an existing code or uses it as a base for the implementation \citep{becciani:2000,becciani:2001,gheller:2002,becciani:2003,payne:2003,comparato:2007,beeson:2003}, while the second trend prefers to build their own custom implementation \citep{gooch:1995,oosterloo:1996,nadeau:2001,magnor:2005}. 

In most of these cases, the emphasis was on volume rendering a dataset that could fit 
entirely in the (local) CPU memory -- appropriate when considering data sizes 
measured in tens of Megabytes (MB) to a few Gigabytes (GB). Handling data cubes larger than 
a single machine's memory limit was first addressed in astronomy by \citet{beeson:2003}, 
using the distributed shear-warp algorithm \citep{lacroute:1994}.  Visualisation 
requirements from other disciplines (e.g. real-time interactive response) has motivated progress in parallel and 
distributed volume rendering solutions for increasingly larger data sizes.

%


\subsection{GPU-cluster volume rendering}

In general, volume rendering techniques can be classified based on the order of data traversal into: image-order [e.g ray-casting \citep{levoy:1988}], object-order [e.g. Splatting \citep{westover:1990}], and hybrid [e.g. Shear-warp factorization \citep{lacroute:1994}].  For details on each of these categories and different volume rendering parallel and distributed implementation trials we refer the reader to \citet{SCHWARZ:2007, molnar:2008}. 
   
The ray-casting process aims to map each pixel on the viewing plane into a colour. The colour and the opacity of each volume element (voxel) is derived from its data value using a predefined mapping operator called a ``transfer function". For each pixel on the viewing plane, the ray-casting process computes a ray originating at this pixel and shoots it into the data volume. By tracing this ray and accumulating the colour and opacity values along the ray, the ray-casting process computes and assigns a final colour and opacity to the pixel. See \citet{levoy:1990} for a detailed description of the original ray-casting algorithm.  

Although it is a computationally intensive task, ray-casting has a simple and clear parallel nature.  This parallel nature has motivated the development of number of parallel ray-casting algorithms with special attention to the usage of GPUs [e.g. \citet{goel:1996}, \citet{scharsach:2005}, \citet{strengert:2006}, \citet{maximo:2008}, \citet{humphreys:2008}, \citet{eilemann:2008}, and \citet{Splotch:2010}]. An extended survey of research on high performance volume rendering using ray-casting and the other alternative rendering approaches can be found in \cite{marmitt:2008}. 

Using a GPU-cluster to perform volume rendering using ray-casting performed in the fragment shader unit and volume bricking was addressed by \cite{muller:2006,muller:2007} .  They investigated the effect of using empty-space skipping, static and dynamic load-balancing approaches, and uniform and non-uniform bricking on the frame rendering time.
\cite{Stuart:2010}  discussed the usage of the MapReduce workflow to implement multi-GPU volume rendering.Their implementation provides both in-core and out-of-core volume rendering based on a CUDA\footnote{\url{http://www.nvidia.com/cuda}} implementation of the ray-casting algorithm.

Many of the previously mentioned trials use OpenGL\footnote{\url{http://www.opengl.org/}} to implement different rendering tasks (e.g. depth sorting) and the implementation of the volume rendering part. That allows them, indirectly, to use GPUs distributed processing capabilities. Within this work, we utilized CUDA to develop our volume rendering algorithm and all the associated rendering tasks. While this made the development task harder, the selection of CUDA gives us two main advantages over OpenGL based distributed rendering. It enables us, relatively more easily, to develop other data analysis and processing tasks that utilize the data already in the GPU memory and the GPU's processing power (e.g. calculating data minimum and maximum values). Also, using CUDA enables our framework to work without the need of XWindows\footnote{\url{http://www.x.org}} or OpenGL which are not supported by some high performance GPUs or due to operational restriction are not easily supported over non-visualization oriented supercomputers (e.g. the previous NVIDIA Tesla Family S1060 and S1070). We think the ability to use general purpose GPU-based supercomputers and clusters (e.g. CSIRO GPU Cluster\footnote{\url{http://www.csiro.au/resources/GPU-cluster.html}} and the gStar Supercomputer\footnote{\url{http://astronomy.swin.edu.au/supercomputing/green2/}}) is an important feature to enable more astronomers to utilize 3D visualization in their day-to-day data analysis and quality control tasks. 


\subsection{An improved solution}
We present an in-core solution for interactive volume rendering datasets that exceed the single 
machine memory limit by using a distributed GPU infrastructure and the ray-casting technique. 
This work represents both enhancement and an extension to our previously published 
work, \citet{hassan:2010a}. 
While the hardware architecture remains as in Figure 2 of \citet{hassan:2010a}, we have 
introduced some significant changes to the software architecture to reduce the communication 
overhead and to speed-up the total rendering time. The main contributions of this work are:
\begin{itemize}
\item Utilize GPU texture memory to speed-up the ray-casting process and to facilitate the usage of tri-linear interpolation. 
\item Investigate the framework's scalability up to 128 GPUs and its ability to render up to 200 GB data cubes.
\item Change the data partitioning to utilize k-dimensional trees (k-d tree) to enable fair data distribution over the contributing nodes.
\item Utilize the rendering rectangle concept further to minimize the communication overhead between different nodes and the server.
\item Utilize asynchronous communication to overlap communication with the merging process.
\item Implement the server merging process over GPU to speed it up and minimize the merging overhead.
\item Dynamic selection of the sampling step to speed-up the ray-casting process while maintaining the rendering accuracy.
\end{itemize}

Overall, these changes enhance the system's scalability, allowing us to make use of new GPU-based
supercomputing clusters and thus interactively visualise even larger data sizes (200 GB compared to 25 GB with our previous solution), and reduce the 
total rendering time, enabling us to achieve frame rates around 30 frames/second (fps) compared to 5 fps with our previous solution.

\section{Distributed GPU Ray-casting Framework}
\label{sc:framework}

In this section, we present the framework's main software components and related design decisions. Also, we elaborate on the new features presented in this work. 

\subsection{Design philosophy}
\label{subsc:basicidea}

The GPU execution model follows a master-slave like execution paradigm. The CPU acts as the main execution controller, controls the access to main memory,
 pushes the data to the GPU local memory,
invokes GPU execution, and pulls the results back to the main memory.  
On the other hand, a GPU is an order of magnitude faster than the CPU in executing single program multiple data (SPMD) kind of operations.  The lack of direct access to the main memory, the limited communication bandwidth between CPU and GPU, and the limited local GPU memory size (currently 6 GB\footnote{\url{http://www.nvidia.com/object/preconfigured-clusters.html}} at maximum) are the main factors restricting the use of GPUs in real-time processing/visualization of larger-than-memory datasets. 
To address some of these issues, the latest generation of GPUs (e.g. NVIDIA Fermi model) are able to communicate and exchange data between each other with a limited CPU involvement.\footnote{\url{http://developer.download.nvidia.com/compute/cuda/4_0/CUDA_Toolkit_4.0_Overview.pdf}} Although such improvements can be effective with GPUs that share the same memory address space (on average 2 GPUs/node), this cannot be easily extended to address the communication between GPUs within different nodes. 

Managing each GPU as a separate instance that synchronizes and exchanges data using the message passing interface (MPI)\footnote{\url{http://www.mpi-forum.org/docs/mpi-20-html/mpi2-report.html}} standard is a straight forward technique to target such architectures.  While the message passing mechanism is generic enough to deal with such situations, the communication overhead caused by the MPI implementation will reduce the system performance. This overhead might not be noticeable while using CPUs, but it is relatively high especially when tens of synchronization and data exchange messages are required per second.

One of the main design objectives of the presented framework is to offer two modes of communication: asynchronous shared memory type of communication between GPUs in the same memory address space; and message passing type of communication between GPUs connected via a network. This combination of shared memory communication and distributed memory communication, and the overlapping between computation and communication, reduces the communication overhead and reduces the needed execution time. 

Separating the process of result display from the back-end compute cluster was done for a practical reason: usually, large GPU-cluster (or high performance computing) facilities do not include appropriate display facilities that can support a high level output resolution or interactivity for the user. Furthermore, some of the high-end high performance computing GPUs do not contain appropriate interfaces for display devices (e.g. NVIDIA Tesla 1060 or 1070 cards). 

To achieve the required volume rendering output each GPU thread is executing the ray-casting algorithm to map a single pixel on the output image into a colour value. 
The selection of ray-casting was based upon the following algorithmic advantages:
\begin{enumerate}
\item The ray-casting algorithm is primarily an image-order volume rendering algorithm, which means its complexity mainly depends on the output image size rather than the input data size. The data size resolution does play a part in determining the number of samples needed per ray, but this has a minor impact as is shown in Section \ref{sc:Results}.
\item The ray-casting algorithm is an embarrassingly parallel algorithm, and is well-matched to the  GPU processing paradigm. 
\end{enumerate}
On the other hand, the disadvantages of using an image-order volume rendering methodology are: 
\begin{enumerate}
\item Image-order volume rendering algorithm requires the whole data to be accessible during the rendering process. This was solved using the bricking technique and proxy geometry [see \citet{lombeyda:2001} for more details].
\item The bricking technique independently maps each sub-volume into the full resolution output frame. Hence, each GPU is required to evaluate the same number of output pixel values. While the proxy geometry is usually used to early-terminate those rays falling outside the current volume, this kind of conditional execution is poorly supported by the GPU architecture and reduces the overall performance. \footnote{\url{http://developer.download.nvidia.com/compute/cuda/4_0_rc2/toolkit/docs/CUDA_C_Programming_Guide.pdf}- chapter 4.} This was addressed by using a CPU-level global termination criteria (rendering rectangle), which minimizes the need for early ray termination on the GPU level. 
\item The main bottle-neck for distributed image-ordered volume rendering methodologies is the compositing operation. We solved this issue by partition the compositing into two-stage compositing, which reduces the communication overhead and the number of tasks that need to be performed by the server node. Also, we utilized rendering rectangles to minimize the number of compositing operations required, and moved the composition step from CPU to GPU.  
\end{enumerate}

\subsection{Hardware and software architecture}
\label{subsc:HWSWArch}

\begin{figure*}
\begin{center}
\includegraphics[scale=0.42]{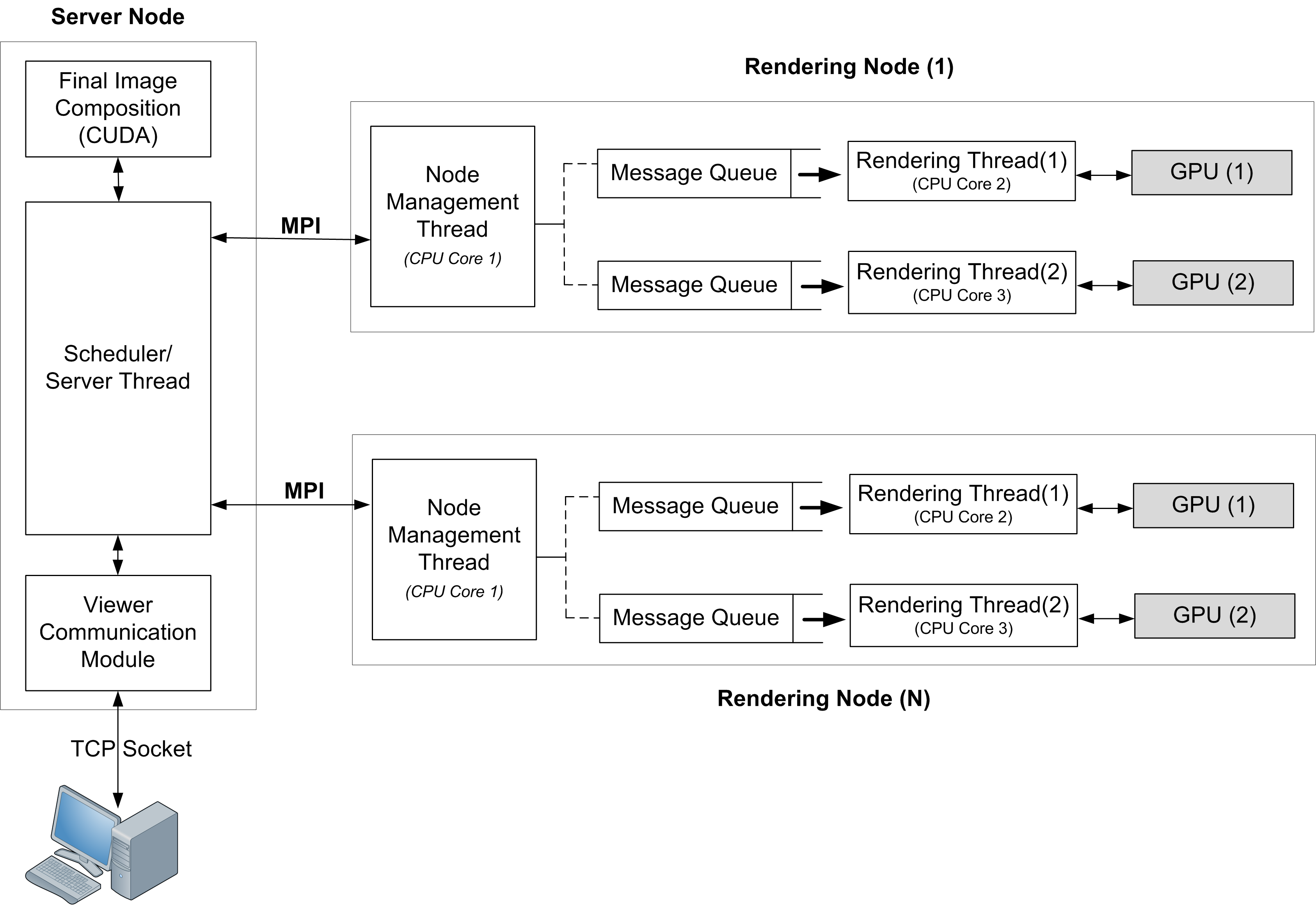}
\caption{Schematic diagram for the main software components of the proposed framework. The communication between the viewer application and the server is done using TCP sockets and is initiated by the viewer application. The communication between the server node and the rendering node is done using MPI. Each GPU is managed and controlled by a separate CPU thread. The internal communication between different GPUs threads within the same workstation is done using a custom message queue and is managed by the root thread.}
\label{fig:schematic}                                 
\end{center}                                 
\end{figure*}

\begin{figure*}
\begin{center}
\includegraphics[scale=0.65]{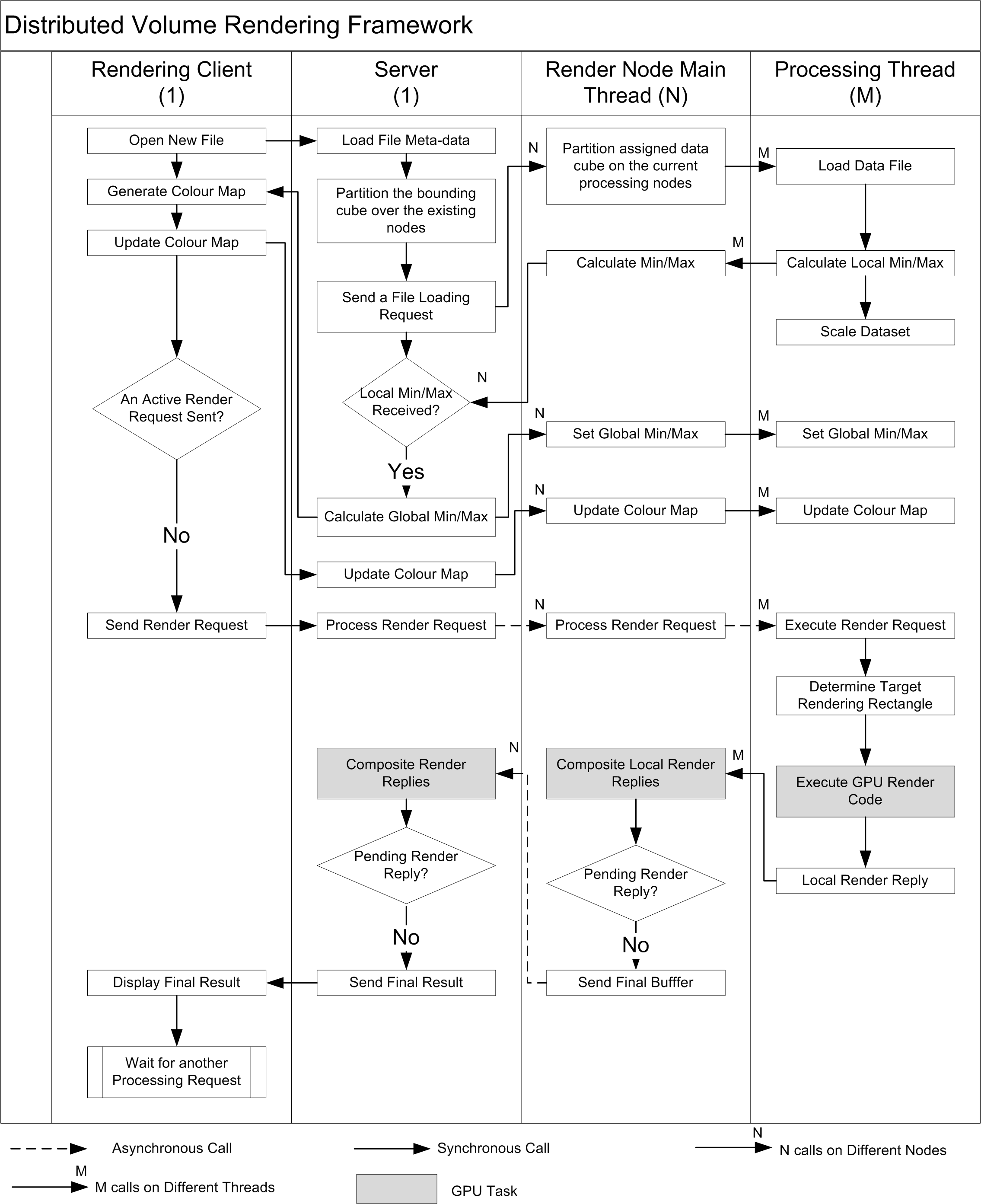}
\caption{Cross Functional Diagram showing the interactions between the system's software components.}
\label{fig:crossfunctional}                                 
\end{center}                                 
\end{figure*}

The main software components of the system are shown in Figure \ref{fig:schematic}. These components are: 
\begin{enumerate}

\item Viewer communication module. This module is based on TCP sockets to enable the server to exchange messages with the viewer machine(s). Messages are exchanged in a custom, predefined binary format. The main task for this module is to interpret the message, identify the required parameters, and notify the server using an event-driven architecture.  This module implements a state machine to identify and validate the possible client communication scenarios. 

\item Server scheduler module and main thread.This module is the main system backbone. It is responsible for partitioning the data cube into sub-cubes, assigning different sub-cubes to rendering nodes and GPUs, synchronizing between different rendering nodes, utilizing the global image composition module to generate the final rendering output, and communicating with the viewer application using the viewer communication module. See Figure \ref{fig:crossfunctional} for details of the different tasks.

\item Global image composition module. This module is mainly a CUDA driver API wrapper, that contains the main functionality required to composite different output sub-frames (received from rendering clients).

\item Rendering node - management  thread. This thread is responsible for initiating and monitoring different rendering threads (one for each GPU in its node), controlling MPI messages and broadcasting them to the rendering threads, compositing the output of the rendering threads, and sending the rendering results back to the server.

\item Rendering node - rendering thread(s). Each of these threads is responsible for managing one GPU card, using the CUDA driver API. These threads are responsible for loading the data and transfering it to GPU texture memory, determining the associated rendering rectangle for each rendering request, and clipping the ray-casting process to it, performing the actual rendering via the GPU, and receiving the output frame back for compositing.

\end{enumerate}

Further details about these modules, and how they communicate, is shown in Figure \ref{fig:crossfunctional} and is discussed in detail in the next section.
For a detailed description of the hardware architecture, we refer the reader to Section 2 and Figure 2 in \citet{hassan:2010a}.

\subsection{Main framework processes}
\label{subsc:frameworkprocesses}
We now describe the framework's main processes and provide a highlight of the details of each software components and how they are integrated together within the framework. 

\subsubsection{Data partitioning and scheduling}
\label{subsubsc:Partitioning}

\begin{figure*}
\begin{center}
\includegraphics[scale=0.6]{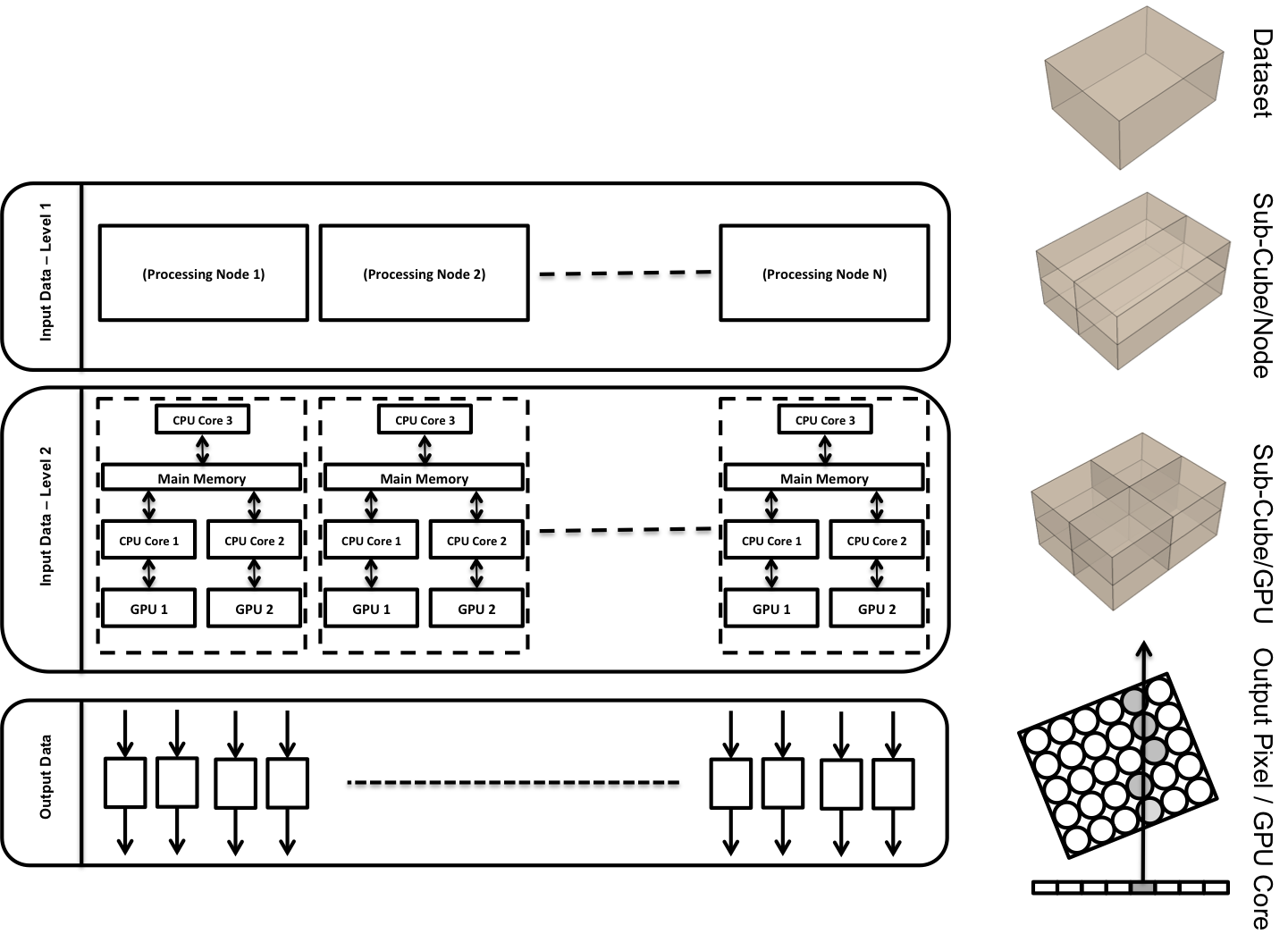}
\caption{Illustration of the data partitioning using k-d tree subdivision and the different levels of granularity. The first level of granularity is the node level (4 nodes in the example). The second level further partitions each node's sub-cube into smaller sub-cubes, which are mapped to the GPUs (8 GPUs with 2 Nodes/GPU). On the GPU level, the resultant image is partitioned where each pixel is allocated to a GPU thread to process. }
\label{fig:datapartitioning}                                 
\end{center}                                 
\end{figure*}

The input data is partitioned into two levels (see Figure \ref{fig:datapartitioning}). The first level partitions the input data cube into smaller sub-cubes based on the number of processing nodes. The size of these sub-cubes should fit in each node's memory, and the total available GPU memory on each node. The second level further partitions each node sub-cube into smaller sub-cubes, which are mapped to the GPUs of each node. The problem is further partitioned by mapping each pixel of the output frame to a GPU thread.

The input-data partitioning process is performed over the server using the file metadata. Based on the file metadata, the input cube is partitioned into a set of uniform sub-cubes using a k-d tree partitioning. The main target of this step is to partition the global data cube into equal  volume partitions, which distributes the required computations fairly over the contributing GPUs. In case the current number of available GPUs does not match the number of nodes required for a balanced k-d tree, the data is partitioned into two or more balanced k-d trees, where the longest axis is used to partition the main cube into the root of each of these k-d trees.  The partition point is selected based on the ratio of the number of GPUs assigned to each k-d tree to the total number of GPUs available. 
The previous partition schema discussed in \citet{hassan:2010a} was not generic enough to achieve a balanced data partitioning over a large number of nodes and GPUs, which is better handled using k-d trees. 
This step aims to minimize the differences between sub-cube dimensions (this contributes to the difference in the frame rendering time with different cube orientation angles, see Section \ref{sc:Results} for details) and to achieve a balanced data assignment between different GPUs. 

Each rendering node sends an independent file loading request(s) to the file server, and starts the file loading process. Whenever the data is loaded in the nodes' CPU memory, the data transformation starts from the CPU memory to the GPU texture memory. An independent task to determine the local minimum and maximum starts when the data is loaded into the GPU memory, and the result is combined with the ``rendering ready'' signal. The output data mapping is updated for each frame based on the proxy-geometry orientation and the computed rendering rectangle.

\subsubsection{Ray-casting process}
\label{subsubsc:raycasting}

\begin{figure*}
\begin{center}
\includegraphics[scale=0.6]{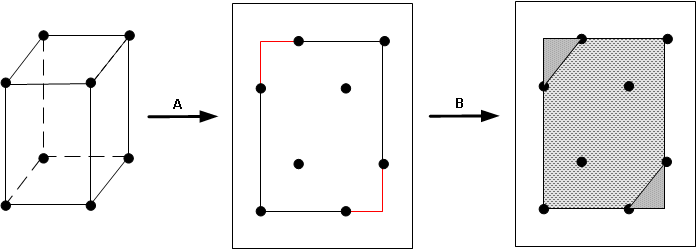}
\caption{Illustration of the process of determining the rendering rectangle for each GPU.
(A) 2D projection of the cube points is calculated, followed by calculating the bounding rectangle.  
(B) The bounding box check is used to exclude the gray region from the output. }
\label{fig:renderingregion}                                 
\end{center}                                 
\end{figure*}

A significant portion of the ray-casting process is executed on GPU using the NVIDIA CUDA driver API. The lack of predictable data access behaviours limits the optimized usage of the global GPU memory and dramatically reduces the data access speed. We choose to store the data cube in the GPU texture memory, which provides the highest available data access speed for such amounts of data. Also, texture memory provides built-in fast linear interpolation functionality. 
The ray-casting process starts on the CPU side with limited pre-processing steps, aiming to speed-up and optimize the ray-casting process.  Using information provided by the viewer application, and passed through the server, about the size of the final output frame, and the OpenGL projection and transformation matrices, each rendering thread projects its data sub-cube vertices onto the output frame.  These projected vertices are used as an input to calculate a bounding rectangle, which contains all the points, to limit the ray-casting process to a specific region of the output frame.   We call this region the ``rendering rectangle'' - see Figure \ref{fig:renderingregion} for an illustration of the process. The rendering parameters are then passed to the GPU kernel,  where a GPU kernel instance is invoked for each pixel within the rendering rectangle. 

Each GPU kernel then starts by calculating its ray's start location, orientation, and where it intersects with the bounding cube. The ray entry and exit point is used to determine a suitable sample step size using a discrete ray-casting with at least one sample for each voxel intersected [see \citet{kaufman:2003} for details]. This provides acceptable output accuracy and processing time especially with the lack of huge variation in the data point values. Also, this check is used to early-terminate any ray lying outside the bounding cube (a few rays may pass the rendering rectangle test but still do not intersect with the bounding cube). The ray-casting process proceeds while the exit point is reached, or the maximum pixel value (defined by the global maximum of the data) or opacity is achieved.  

\subsubsection{Final image composition}

The image compositing process is performed over two layers. The first layer is on each of the processing nodes that contains more than one GPU. Using a shared rendering buffer, initially blank, each GPU projects its rendering data over this buffer with the guidance of its rendering rectangle. The compositing process is achieved using a GPU kernel, and the rendering rectangles are merged together to form the whole-node rendering rectangle.  The whole-node rendering rectangle (a union of the rendering rectangles of each GPU) is then used to clip the rendering buffer, and only the area within the rendering rectangle is sent to the server along with the rendering rectangle information.

The server receives the individual rendering sub-buffers into a separate queue to achieve best overlapping between the communication and the process of server merging. Using each node rendering rectangle, the server merges the node results with the final rendering buffer. The server merging process is also performed over GPU. When all the received sub-buffers are merged, the server initiates a colour mapping process using the user specified colour map and sends the results back to the client.   

The  composition complexity and validity depends on the selected transfer function. The currently adopted transfer function is the maximum intensity projection (MIP)\citep{Wallis:1989}. It was selected because of its straight-forward, easy to interpret, mapping between the data and the used colour map,  ability to emphasis features with local maximum (e.g. large scale noise patterns and Radio Frequency Interference), and low parallelization overhead. The usage of MIP as the main transfer function allows arbitrary compositing order which facilitates and speeds-up the compositing process. \citet{lombeyda:2001} provide a mathematical proof that the general alpha-blending volume rendering operator is associative which allows performing the compositing step in parallel.

\subsubsection{Communication}
\label{subsubsc:Comm}

The communication overhead is the main bottleneck for the whole system's scalability. After trying different methodologies, including multi-layer compositing [see \citet{beeson:2003}], using different dedicated compositing nodes, and even applying different compression algorithms, the direct send methodology was found to be the best from the perspective of the total rendering time and the cost of the hardware needed. The same conclusion was reached by \citet{Stuart:2010}, \citet{muller:2006}, and \citet{muller:2007}. 

The average rendering result message size, without compression or encoding, is 4 MB for a 1024 x 1024 output frame.  This amount of data increases linearly with the number of nodes in this case, which means that adding new GPUs does not necessarily decrease the total frame rendering. 
Here, using the individual node render rectangle to determine the amount of non-blank data, the amount of data transferred between the nodes and the server is no longer constant. 

Based on the rendering rectangles of individual nodes, each node rendering buffer is packed in a smaller message, whose size depends on the cube orientation and the amount of data assigned to each node.  Furthermore, each node utilizes its local memory to do the first local composition step which further reduces the amount of data to be transferred to the server. With the message packing mechanism and the two stage compositing the total amount of data transferred for each frame is reduced from $N \times M$ to $M+\epsilon$ , where $N$ is the number of rendering nodes, $M$ is the final frame size in bytes, and $\epsilon$ is a slight increase in the size caused by the overlapping between different rendering rectangles. This reduction remove a potential bottle neck for the framework scalability and significantly reduces the total rendering time. Also, the usage of the rendering rectangles speeds up the server compositing process, and allows better overlapping between communication and computations because it excludes applying the compositing operator over non-overlapped pixels.

\section{Results and discussion}
\label{sc:Results}
\subsection{Performance analysis and timing tests}
\label{subsc:timing}

Performance analysis and timing test were performed on the Australian Commonwealth Scientific and Industrial Research Organisation (CSIRO) GPU cluster. This GPU cluster contains 128 nodes with two GPUs each. The nodes are identical Dual Xeon E5462 compute nodes connected via a DDR InfiniBand Switch. The timing tests were performed using both NVIDIA Tesla 1060 and Tesla 2050\footnote{\url{http://www.nvidia.com/docs/IO/56483/Tesla_C1060_boardSpec_v03.pdf}} GPUs.  Table \ref{tab:ListOfDatasets} shows the details of the datasets used to evaluate the framework performance.   The tests were performed with different numbers of nodes ranging from 2 to 64 nodes (each with two GPUs) with the data size limiting the minimum number of nodes used. Our test cases include different astronomical data types including dark matter simulation data [smoothed over a structured grid using cloud-in-cell technique \citep{hockney:1988} - to show the applicability of our approach to visualize other astronomical data types] and spectral data cubes from different astronomical surveys.

\begin{figure*}
\begin{center}
\includegraphics[scale=0.7]{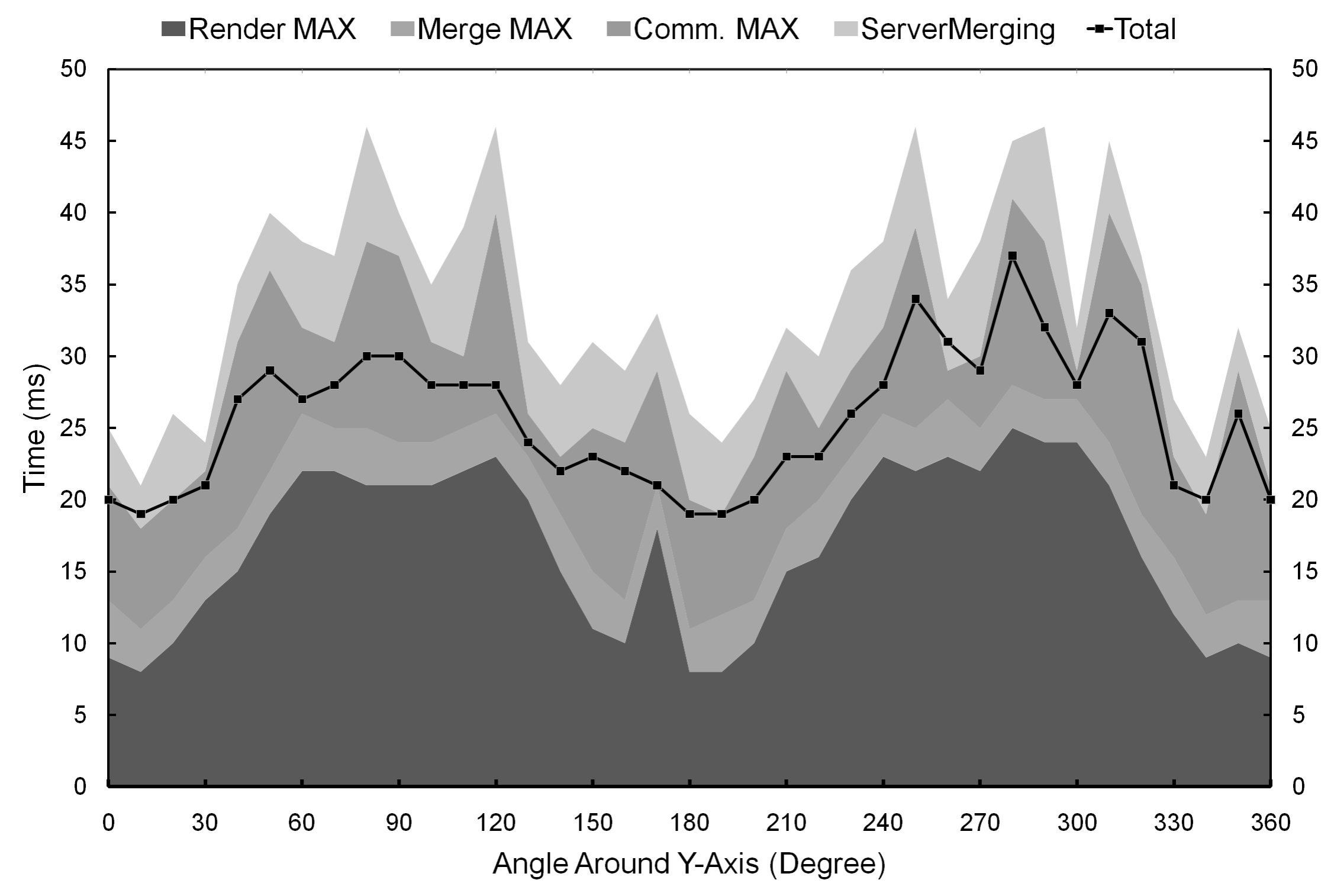}
\caption{Single frame rendering time for different cube rotation angles. The timing measurements were done for the Scaled  GASS Cube (203.8 GB) on 65 workstations (one acts as a server) and 128 GPUs.}
\label{fig:GASS204DetailedTiming}                                 
\end{center}                                 
\end{figure*}

Due to the overlapping between communication and computation and the shared memory access synchronization, it is hard to find an exact equation to govern the total frame rendering time in terms of its sub-rendering processes. 
Total frame rendering time (TR) is directly related to the maximum GPU rendering time,  the maximum local compositing time, the maximum communication time, and the total server compositing time. 
Figure \ref{fig:GASS204DetailedTiming} shows the relation for the scaled GASS dataset rendered with 64 nodes (128 GPUs), with the measured total frame rendering time displayed as a function of cube orientation. The different communication patterns and the overall cluster usage (e.g. from other applications) during our tests may slightly affect the communication delays and so the final rendering time of each frame. To reduce the impact of such effects, each data point is the median of 10 rendering runs with the same rotation angle. 
It is clear that the GPU rendering time is the leading factor, varying with the cube orientation (defined in the term of the rotation angle around the Y-axis). This change in the data cube orientation affects the size of the rendering rectangles, the amount of overlapping between them, the length of the rays cast, and the number of samples calculated for each of the rays. Due to this variation, we use the 75\% percentile to represent the estimated frame rendering time for each data cube.  

\begin{figure*}
\begin{center}
\includegraphics[scale=0.7]{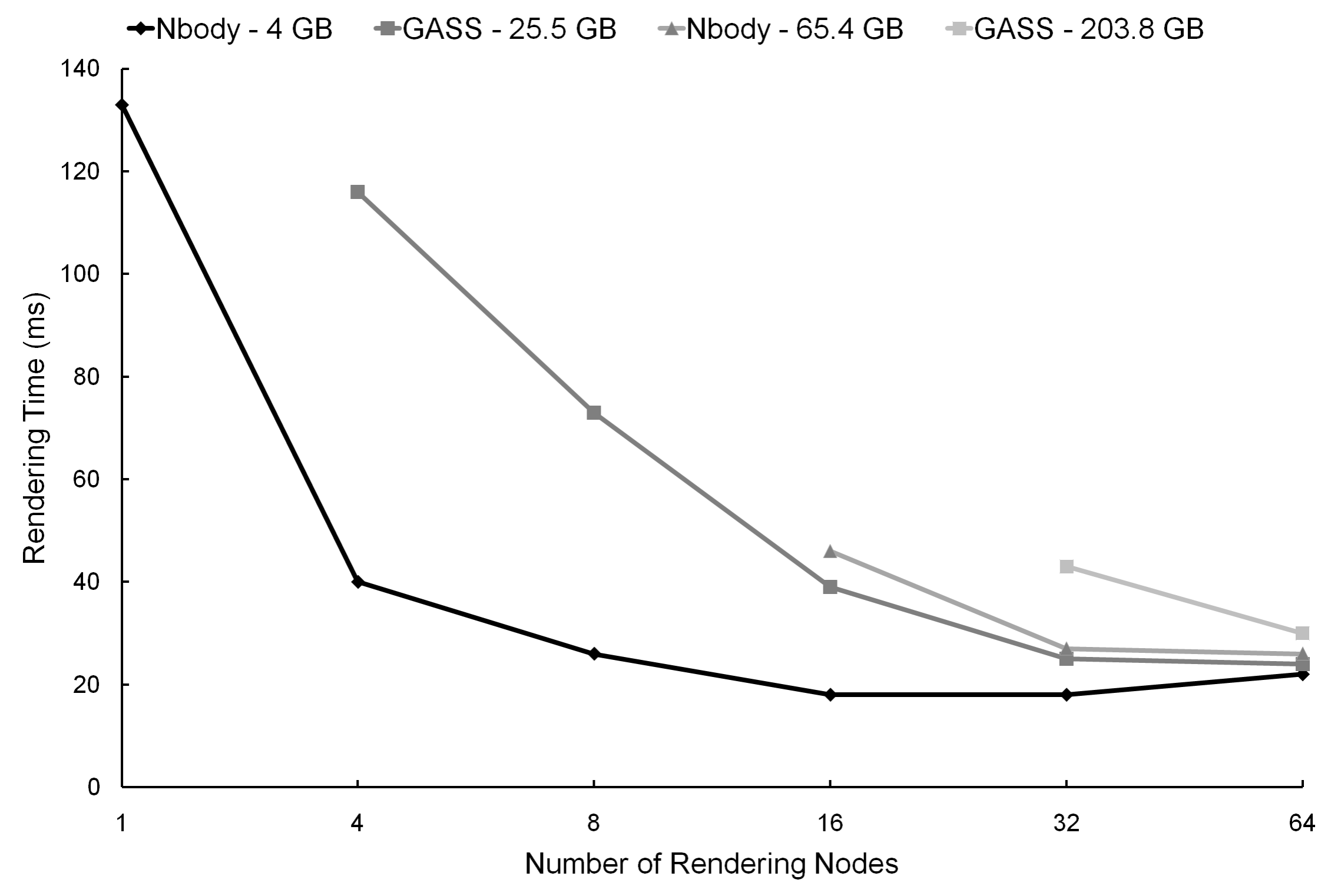}
\caption{Frame rendering times for the different test cases with different numbers of rendering nodes, using Tesla 1060 GPUs and an output frame size of 1024 $\times$ 1024.}
\label{fig:AllTiming}                                 
\end{center}                                 
\end{figure*}

Figure \ref{fig:AllTiming} shows a summary of the timing tests performed using all the test cubes over number of nodes from 2 to 64, with an output frame size of 1024 $\times$ 1024, for the Tesla 1060 GPUs. In general, the total frame rendering time is reduced by the introduction of new GPU nodes, however, the effect of this reduction reaches a critical point where the increase in the GPU nodes slightly increases the total rendering time or else appears to keep it constant. At this critical point, usually the time spent in the ray-casting and merging is much lower than the communication overhead, because of the tiny size of the problem (e.g. sub-data cubes occupy less than 1\% of the GPU memory in the case of the 4 GB Nbody simulation cube with 32 nodes).    

\begin{figure*}
\begin{center}
\includegraphics[scale=0.65]{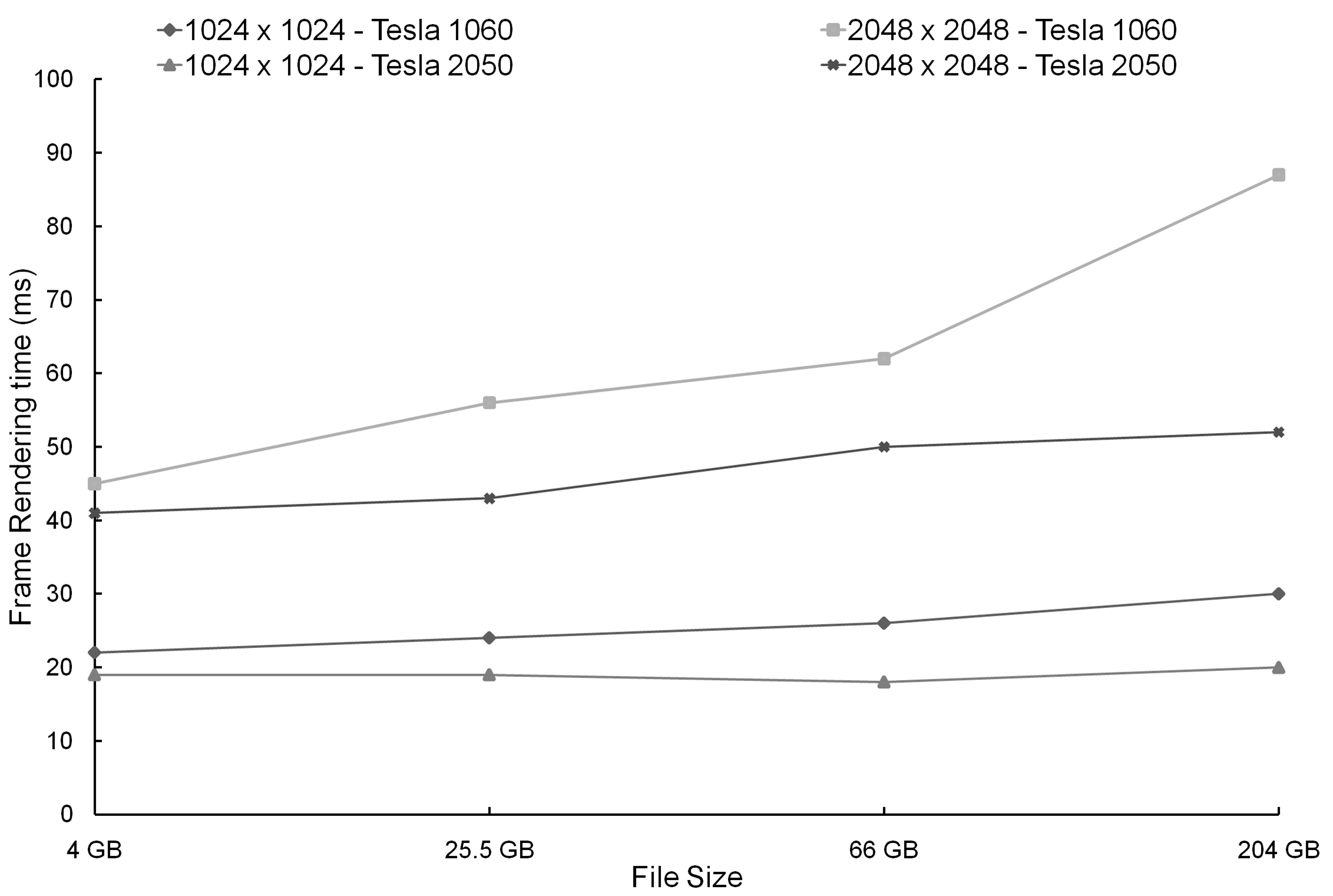}
\caption{The effect of increasing the output frame resolution on the rendering time from 1024 $\times$ 1024 to 2048 $\times$ 2048 for different test cases on 64 rendering nodes with Tesla 1060 and 2050 GPUs .}
\label{fig:differentOutput}                                 
\end{center}                                 
\end{figure*}

Figure \ref{fig:differentOutput} shows the effect of increasing the output frame size from 1024 $\times$ 1024 to 2048 $\times$ 2048 on the total rendering time for different test cubes with 64 nodes and Tesla 1060 as the GPU unit, and shows the same test with a comparison between the timing on Tesla 1060 and 2050. The output frame size is the major factor in determining the number of rays in the ray-casting algorithm, and affects the data size communicated in each frame. The usage of Tesla 2050 instead of Tesla 1060 (with an increase around 180\% in the number of GPU cores but with lower memory size and core frequency) gives a reduction in the total rendering time by 35\% on average (ranging from 10\% to 67\%) .

\begin{table*}
\caption{Sample datasets used to evalute the performance of our framework.}
	\label{tab:ListOfDatasets}
	\centering
		\begin{tabular}{ccp{5cm}c}
		\hline
			Dataset Name & Dimensions (Data Points) & Source / Credits & File Size\\ \hline	
			Nbody cube  & 1024 x 1024 x 1024 & High resolution $1080^3$ dark matter simulation of a 125 Mpc/h box by Swinburne Computations for WiggleZ (SCWiggleZ) project (Poole et al 2010, in prep)  & 4 Gigabyte \\ \hline								
			HIPASS Cube & 1721 x 1721 x 1025 & HIPASS Southern Sky, data courtesy Russell Jurek HIPASS team & 12 Gigabyte  \\ \hline
			GASS Cube & 2502 x 2501 x 1093 & The Parkes Galactic All-Sky Survey, data courtesy Naomi McClure-Griffiths/ GASS team \citep{Griffiths:2009} & 25 Gigabyte  \\ \hline
			Scaled Nbody cube  & 2600 x 2600 x 2600 & Scaled version of the Nbody cube  &  65.4 Gigabyte \\ \hline
			Scaled GASS Cube & 5004 x 5002 x 2186 & Scaled version of the GASS cube & 203.8 Gigabyte  \\ \hline
		\end{tabular}	
\end{table*}

\subsection{Discussion}
\label{subsc:Discussion}
We performed the scalability and performance tests for the proposed framework for different test cases, different GPU architectures, and different output resolutions. A frame rate larger than 10 fps is achievable even with an output resolution exceeding the standard 1024 $\times$ 1024 of typical desktop monitor. The system scales with an increase in the GPU count till a certain critical point where the problem is too tiny for the GPUs to solve: in such cases the communication time exceeds the GPU rendering and compositing time, and may cause a slight increase in the total frame rendering time. 
The proposed system enables us to render datasets beyond the current single machine memory limit with a frame rate up to 30 fps for an output size of 1024 $\times$ 1024 and gives promising results for higher output sizes of 2048 $\times$ 2048 (despite the current  lack of standard desktop displays capable of showing such an image at full resolution). 

The main disadvantage of this system is the need for GPU memory that matches the amount of data to be rendered. We think the in-core solution is the best available alternative to achieve a real-time volume rendering and high level of interactivity for terabyte-order  datasets. The huge dataset size, especially with the hard disk I/O speed as a limiting factor, is very hard to be rendered using out-of-core methodologies. Also, the use of a multi-resolution solution is not practical due to the high dynamic range inherent in many astronomical spectral data cubes, the lack of well-defined object boundaries, and the low-signal to noise data values.  Furthermore, introducing interactive quantitative visualization along with the qualitative visualization, which is one of our main future-work goals, needs the whole dataset to be available in memory to offer a near real-time response for the users’ queries. We expect the memory limiting condition to be relaxed during the next years with advances in GPU architectures and the increase in available memory per GPU.  Most critically, from a financial and cluster management perspective, the number of GPUs needed to render a certain dataset is much fewer than the number of CPU cores needed to render the same dataset. 

Using data migration or dynamic load balancing may not be effective in our case; because of the memory limitation that enforces the maximum amount of data that can be stored in the GPU memory. Also, the communication overhead with such data migration may slow down the rendering operation, especially when a high resolution output is required. Static load balancing may be useful to reduce the variation of the frame rendering time between different orientation angles. The overlapping between communication and computation will reduce the effect of such load balancing. The server has no expectation for a certain node to be faster and there is no direct relation between that and the final rendering time.  A near cubic data partitioning schema (where the ray lengths and the rendering rectangle areas are almost the same with different cube orientation) minimizes the differences between rendering time for each orientation angle.  We leave further static load balancing investigations as a future work. 

\subsection{Future Work}
\label{sc:future}

Changing the viewer application into a web-based application using Java3D, FLASH, or HTML5 is one of the steps to enable wider usage of remote visualization in astronomy. Such platform changes are easily integrated with the current framework.  Also, enabling quantitative data visualization techniques with more interactivity and control given to the user is an important addition to enable better scientific results and improved knowledge discovery. Additionally, more work is needed to develop a customized astronomical data transfer function which can suppress the noise and enable discoveries in the low-signal to noise values.

\section{Conclusion}
\label{sc:conc}
We presented an enhanced framework to volume render larger-than-memory spectral data cubes. The framework utilizes a hybrid infrastructure of shared- and distributed-memory high performance computing architectures to enable interactive rendering of datasets that exceed single machine memory limits. The framework utilizes an optimized version of the ray-casting algorithm and the volume bricking data partitioning mechanism to distribute the volume rendering task over a cluster of GPU-powered workstations connected over a high-speed network. Using available knowledge about the cube orientation and how the data is partitioned over the nodes, the framework optimizes the amount of data that needs to be transferred between the different contributing nodes in order to improve the scalability and reduce the frame rendering time. Using two stage compositing reduces the amount of work needed by the server and reduces the total amount of data needed to be transferred to prepare each frame. Using GPUs as the main processing element for the rendering and compositing processes reduces the total rendering time and removes the compositing bottle neck. Moreover, using GPUs reduces the number of nodes needed, the cost needed to visualize such large data cubes, and the communication overhead required. 

The framework enables remote visualization by separating the actual results display from the rendering computations. This remote visualization facility can be used to enable a wider usage of the service for geographically distributed users without the need for tedious moving of data between collaborators. 
Also, this separation enables the usage of different viewing platforms including the Web. Different timing analyses were made to test the framework performance and scalability. The framework was able to render data cubes up to 204 GB in size at 30 frames per second using 128 GPUs. Our timing tests also investigated the effect of increasing the output resolution and of upgrading the GPUs from Tesla 1060 to Tesla 2050. We anticipate, based on these results, that the framework can continue its scalability with a little increase in the communication overhead to render larger data cubes with a comparable performance.

Although our focus has been on volume rendering spectral data cubes from radio astronomy, 
we contend that our approach has applicability to other disciplines where  
close to real-time volume rendering of terabyte-order 3D data sets is a requirement. For astronomers, the good news is that real-time interactive visualization of Terascale spectral data cubes is within reach.

\section*{Acknowledgments} 
We thank Dr. Tim Cornwell (ATNF) and Dr. John A Taylor (CSIRO) for providing us with an access to the CSIRO GPU cluster.
We thank Dr. Virginia Kilborn, Dr. Emma Ryan-Weber, and Dr. Gregory Poole (Swinburne University of Technology), Dr. Russell Jurek and Dr. Naomi McClure-Griffiths  (ATNF - CSIRO), and Dr. Tara Murphy (Sydney University) for providing sample data cubes, useful discussions, and suggestions.

\bibliographystyle{elsarticle-harv}        
\bibliography{references}



\end{document}